\documentclass[aps,twocolumn,superscriptaddress]{revtex4}
\usepackage{graphicx}

\begin{document}

\title{Exciton-relaxation dynamics in lead halides}

\author{Masanobu Iwanaga}
\email[Electronic address: ]{iwanaga@phys.h.kyoto-u.ac.jp}
\affiliation{Graduate School of Human and Environmental Studies, Kyoto 
University, Kyoto 606-8501, Japan}
\author{Tetsusuke Hayashi}
\affiliation{Department of Fundamental Sciences, Faculty of Integrated 
Human Studies, Kyoto University, Kyoto 606-8501, Japan}

\date{\today}

\begin{abstract}
We survey recent comprehensive studies of exciton relaxation in the crystals 
of lead halides. The luminescence and electron-spin-resonance studies have 
revealed that excitons in lead bromide 
spontaneously dissociate and both electrons and holes get self-trapped 
individually. Similar relaxation has been also clarified in lead 
chloride. The electron-hole separation is ascribed to repulsive 
correlation via acoustic phonons. Besides, on the basis of the temperature 
profiles of self-trapped states, we discuss the origin of 
luminescence components which are mainly induced under one-photon excitation 
into the exciton band in lead fluoride, lead chloride, and lead bromide. 
\end{abstract}

% insert suggested PACS numbers in braces on next line
%\pacs{}
% insert suggested keywords - APS authors don't need to do this

\maketitle

% body of paper
\section{Introduction}
Most of lead-compound crystals have the bandgap in connection with the 
$6s$-to-$6p$ gap in lead ions and tend to become highly luminescent coming 
from the `odd' transition. 
In fact, excitonic transitions in lead halides are partly explained by 
the $6s$-to-$6p$ transition in lead ions \cite{Liidja,Fujita}. 
Luminescence in cubic PbF$_2$, PbCl$_2$, 
and PbBr$_2$ is composed of broad Gaussian bands with large-Stokes shifts 
and is indicative of strong 
exciton--acoustic-phonon interaction \cite{Liidja} 
while luminescence in PbI$_2$ shows free-exciton (FE) natures \cite{Kleim}. 
At the early stage of the study, the luminescence in lead halides was 
attributed to electric dipole transition from the $6p$ to $6s$ states 
\cite{Liidja,Rybalka}. 
However, the description is too simple to explain the experimental results 
such as two-photon excitation spectra of photoluminescence (PL) 
\cite{Iwanaga1,Iwanaga3}. Cluster calculation for cubic PbF$_2$, PbCl$_2$, and 
PbBr$_2$ shows that the conduction bands consist of Pb$^{2+}$ ($6p$ states), 
and the top of valence bands is composed of 68\% Pb$^{2+}$ ($6s$ states) and 
32\% F$^-$ ($2p$ states), 48\% Pb$^{2+}$ ($6s$) and 52\% Cl$^-$ ($3p$), 
and 35\% Pb$^{2+}$ ($6s$) and 65\% Br$^-$ ($4p$), respectively 
\cite{Fujita}. The tendency of the mixed ratio in the valence bands is 
qualitatively supported by the recent 
electron-spin-resonance (ESR) study manifesting self-trapping hole centers 
of Pb$^{3+}$ in PbCl$_2$ \cite{Iwanaga4} and Br$_2$$^-$ in PbBr$_2$ 
\cite{Iwanaga2}. 

Mixed crystal of PbI$_{\rm 2(1-x)}$Br$_{\rm 2x}$ is an example to show 
the variety of exciton dynamics coming from lead-halogen complex; it enables 
us to observe the change from FE luminescence into self-trapped-exciton 
(STE) luminescence 
by increasing the ratio of bromide 
ions \cite{Takeda}. This system implies that halogen ions add the variety 
in exciton dynamics and play a crucial role in realizing the localization 
of excitons. This example tells us that the simple description based on the 
inner transition of lead ions is imperfect in discussing the lattice 
dynamics of excitons. 

Moreover, it was recently reported that composite 
luminescence in CsPbCl$_3$, comprised of FE and 
STE luminescence, presents the recovery of 
FE-luminescence intensity and the sudden disappearance of the STE luminescence 
at a phase transition under the condition of raising temperature 
\cite{Hayashi}; the excitonic transition stems from 
octahedrons of Pb$^{2+}$(Cl$^-$)$_6$, and the crystal has 
the valence and conduction bands similar to PbCl$_2$ \cite{Heidrich}. 
Thus, lead-halogen complex makes it possible to realize a diversity of 
exciton relaxation. 

We overview recent experimental results on PbBr$_2$ and PbCl$_2$ in the 
next section, and discuss the exciton relaxation resulting in spontaneous 
electron-hole separation in Sec.\ \ref{discussion}. 
Furthermore, the properties of `STE-like' luminescence in cubic PbF$_2$, 
PbCl$_2$, and PbBr$_2$ are examined from comparison with electronic localized 
states. 

\begin{figure}[b]
\begin{center}
\includegraphics[height=56mm,width=70mm]{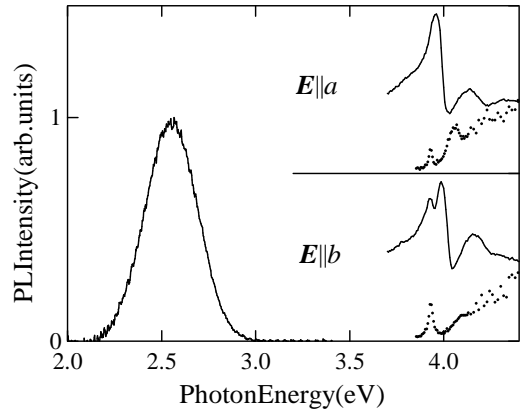}
\caption{%
Photoluminescence (PL) spectrum (left solid line), polarized 
reflectance spectra (right solid line), and two-photon excitation spectra 
(dots) of PbBr$_2$. All spectra were measured below 8 K. The PL was induced 
by 4.43-eV photons.}
\label{fig1}
\end{center}
\end{figure}

\section{Experimental results: lead bromide and lead chloride}
We mainly show optical and luminescent properties of PbBr$_2$ in this section 
because PbBr$_2$ and PbCl$_2$ are similar in the electronic-band structures 
\cite{Fujita} and luminescence properties \cite{Liidja,Iwanaga3,Kitaura}. 
Localized states of electrons and holes in PbBr$_2$ and PbCl$_2$ 
are also presented, which have been investigated with the 
ESR technique. More details of the luminescence 
spectroscopy and ESR study have already been reported in Refs.\ 
\cite{Iwanaga1,Iwanaga3,Iwanaga4,Iwanaga2}. 

\begin{figure}[t]
\begin{center}
\includegraphics[height=100.6mm,width=70mm]{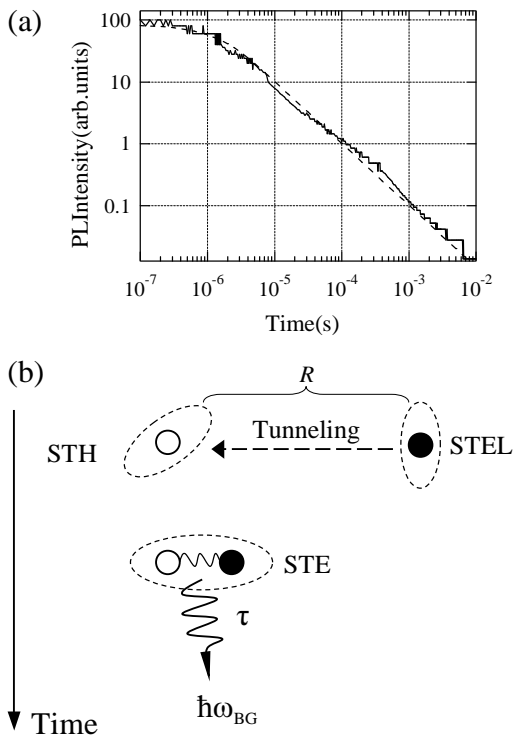}
\caption{%
(a) Decay curve (solid line) of PL in Fig.\ \ref{fig1}. (b) 
Recombination model that reproduces the decay curve; the calculated curve 
from the model is the dashed line in (a).}
\label{fig2}
\end{center}
\end{figure}

Figure \ref{fig1} shows photoluminescence (PL) spectrum (left solid line), 
reflectance spectra (right solid line) for $\mathbf{E}$$\parallel$\textit{a} 
and $\mathbf{E}$$\parallel$\textit{b} configurations, and two-photon 
excitation spectra (dots) of PbBr$_2$; 
all spectra were measured below 8 K. The PL at 2.55 eV was induced 
by 4.43-eV photons and is called blue-green-PL (BG-PL) band from now on. 
The two-photon excitation spectra indicate discrete peaks of exciton 
absorption at 3.93 and 4.07 eV, and the band edge is identified at 4.10 eV 
from the continuous rise. Consequently, the binding energy of the lowest 
exciton is estimated to be 170 meV. Similarly, the binding energy in PbCl$_2$ 
is found to be 200 meV \cite{Iwanaga3}. 

Figure \ref{fig2}(a) presents a decay curve (solid line) of the BG-PL band 
in Fig.\ \ref{fig1}; the curve is plotted with the log-log scale. 
The intensity decays in proportion to $1/t$ for $t\ge$10 $\mu$s. 
Such a decay curve cannot be explained by radiative transition in two-level 
systems but requires a process including tunneling motion \cite{Delbecq}. 
Figure \ref{fig2}(b) depicts an electron-hole recombination model 
for the BG-PL 
band; a pair of a self-trapped electron (STEL) and a self-trapped hole (STH) 
separated by distance $R$ gets close by tunneling motion, forms a STE, 
and recombines with the decay time $\tau$. The $\tau$ was determined by 
time-resolved PL measurement in the microsecond range and is equal to 1.2 
$\mu$s \cite{Iwanaga1}. By assuming the distribution of pair density 
$n(R)\propto R^{-2}$, the model reproduces the measured decay curve well; 
the calculated curve is expressed as $I(t) = (A/t)[1-\exp(-t/\tau)]$, where 
$A$ is a proportionality constant, and is represented with the dashed line 
in Fig.\ \ref{fig2}(a). In PbCl$_2$, an intrinsic BG-PL band at 2.50 eV also 
shows a phosphorescent decay profile 
similar to the BG-PL band in PbBr$_2$ \cite{Iwanaga3}. 

\begin{figure}[t]
\begin{center}
\includegraphics[height=51.1mm,width=70mm]{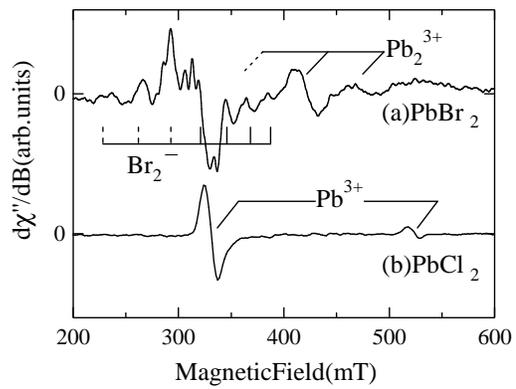}
\caption{%
Typical ESR spectra of (a) PbBr$_2$ and (b) PbCl$_2$
after photoirradiation below 8 K with 120-fs-width and 3.10-eV laser pulses. 
Both spectra were measured below 8 K. The resonant microwave frequencies 
were (a) 9.385 GHz and (b) 9.400 GHz. Solid lines indicate resonances, and 
dashed lines show the ESR positions calculated from the spin Hamiltonian in 
Ref.\ \cite{Iwanaga2}.}
\label{fig3}
\end{center}
\end{figure}

The recombination model hypothesizes the existence of STELs and STHs in 
PbBr$_2$. In fact, they are found in the ESR measurements as shown in Fig.\ 
\ref{fig3}; the ESR signals measured below 8 K were induced by 
photoirradiation below 8 K with pulsed 120-fs-width light at 3.10 eV, 
which induces two-photon interband transitions efficiently. 
Curve (a) in Fig.\ \ref{fig3} presents the STEL centers of Pb$_2$$^{3+}$ 
and the STH centers of Br$_2$$^-$ in PbBr$_2$. The configurations of the 
STEL and STH centers were determined from the spin-Hamiltonian analysis of 
rotation-angle dependence of the ESR signals \cite{Iwanaga2}; 
the dimer-molecular STEL centers orient along the $a$ axis, 
and, on the other hand, the STH centers have two possible 
configurations in the unit cell, which are symmetric for the $bc$ plane, 
reflecting the crystallographic symmetry. 
Experimental finding of 
coexistence of STELs and STHs is the first case to our knowledge. 

Curve (b) in Fig.\ \ref{fig3}, which was measured like curve (a), 
shows the STH centers of Pb$^{3+}$ in PbCl$_2$. Though the STEL 
centers of Pb$_2$$^{3+}$ in PbCl$_2$ were first observed after 
x- or $\gamma$-ray irradiation at 80 K \cite{Nistor,Hirota}, 
the STEL centers are not detected at 0--1700 mT after photoirradiation 
below 50 K as shown in Fig.\ \ref{fig3}(b); 
indeed, the STEL centers of Pb$_2$$^{3+}$ 
appears thermally above 60 K \cite{Iwanaga4}. 
If electrons below 50 K do form self-trapping centers of Pb$^+$ or 
Pb$_2$$^{3+}$, they could be detected in the range of 0--1700 mT 
because they usually have $g$ values of 1--1.6 \cite{Nistor,Hirota,Goovaerts}. 
However, since they are not observed, other electron-trapping centers 
associated with lattice defects are most likely below 50 K. 

Thermal stability of the STELs and STHs in PbBr$_2$ supports the correlation 
with the BG-PL band; in particular, the quenching of the STHs at 20--30 K 
corresponds to that of the BG-PL band. Thus, the STEs responsible for the 
BG-PL band are closely associated with the STHs. 

\begin{figure}[t]
\begin{center}
\includegraphics[height=81.7mm,width=65mm]{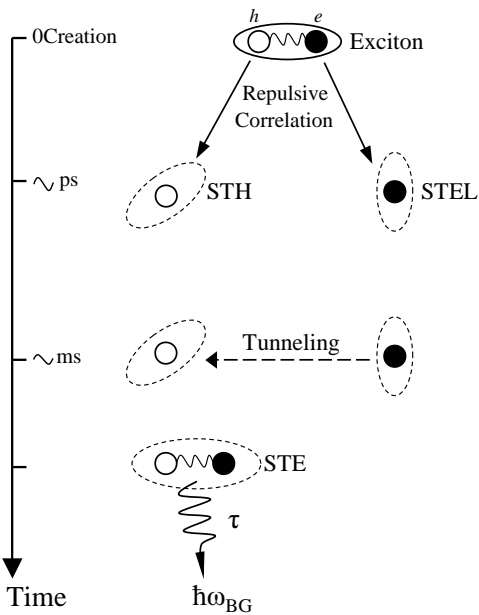}
\caption{%
Schematic description of exciton-relaxation dynamics in PbBr$_2$ 
and PbCl$_2$. In the latter crystal, the STEL has to be replaced with a 
trapped electron. }
\label{fig4}
\end{center}
\end{figure}

The BG-PL band observed  below 100 K in PbCl$_2$ corresponds to the STH 
centers of Pb$^{3+}$ in the stable temperature range \cite{Iwanaga4}. 
Therefore, the recombination model in Fig.\ \ref{fig2}(b) is applicable to 
the BG-PL band in PbCl$_2$ except for replacing the STEL with a trapped 
electron.

\section{Exciton relaxation in lead halides\label{discussion}}
Phosphorescent decay of the BG-PL bands in PbBr$_2$ and PbCl$_2$ is closely 
related to spatially separated electron-hole pairs as depicted in 
Fig.\ \ref{fig2}(b). 
However, can it happen that excitons with the binding energy of about 200 meV 
dissociate spontaneously?

Relaxed states of excitons were theoretically classified by Sumi \cite{Sumi}; 
as a result, excitons can relax into pairs of a STEL and a STH when both 
electrons and holes strongly interact with acoustic phonons. 
Particularly, when the signs of coupling constants of 
electron--acoustic-phonon and hole--acoustic-phonon interactions 
are opposite, repulsive force interplays between the electron and hole 
via acoustic phonons. Thus, the repulsive correlation is ascribable to the 
origin of electron-hole separation. Taking all the results and discussion 
into account, exciton-relaxation dynamics in PbBr$_2$ and PbCl$_2$ is 
schematically depicted in Fig.\ \ref{fig4}. 
The spontaneous breaking of initial bound states, namely excitons, 
takes place by the repulsive correlation via acoustic phonons; 
in view of bipolaron dynamics, the breaking is in contrast with the formation 
of Cooper pairs mediated by acoustic phonons. 

Table \ref{lead_halides} summarizes the self-trapped states of excited 
electrons and holes. Self-trapping is observed in cubic PbF$_2$, PbCl$_2$, 
and PbBr$_2$; in particular, self-trapping of both electrons and holes 
is realized in PbCl$_2$ and PbBr$_2$. STHs in cubic PbF$_2$ irradiated with 
$\gamma$ rays or neutrons were observed only at 77 K, and 
STELs were not detected at the temperature \cite{Nishi} though 
they might be observed at temperatures lower than 77 K. 

\begin{table}[t]
  \begin{center}
    \begin{tabular}{c|cc}
    \hline
    \hline
    Crystals & STEL & STH \\
    \hline
    Cubic PbF$_2$ & $\times$ & Pb$^{3+}$ \cite{Nishi}\\
    PbCl$_2$ & Pb$_2$$^{3+}$ \cite{Nistor,Hirota}& Pb$^{3+}$ \cite{Iwanaga4}\\
    PbBr$_2$ & Pb$_2$$^{3+}$ \cite{Iwanaga2}& Br$_2$$^-$ \cite{Iwanaga2}\\
    PbI$_2$ & $\times$ & $\times$ \\
    \hline
    \hline
    \end{tabular}
  \end{center}
  \caption{
  Structures of STEL and STH centers in lead halides. $\times$ denotes 
  no report of self-trapping.}
  \label{lead_halides}
\end{table}

\begin{figure}[b]
\begin{center}
\includegraphics[height=45.7mm,width=70mm]{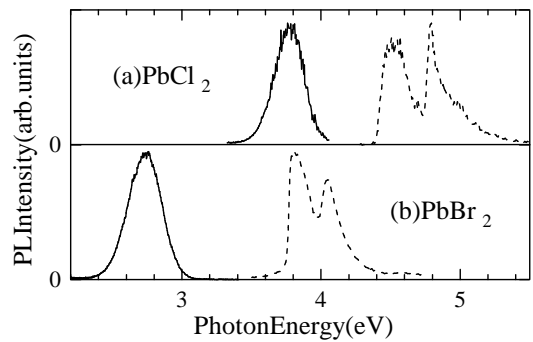}
\caption{%
(a) PbCl$_2$ and (b) PbBr$_2$: PL (solid lines) and the excitation spectra 
(dashed lines). All spectra were measured below 8 K. PL in (a) and (b) was 
induced by 4.80-eV and 3.81-eV photoexcitation, respectively. The excitation 
spectra were observed at (a) 3.76 eV and (b) 2.95 eV.}
\label{fig5}
\end{center}
\end{figure}

From comparison with the stable temperature range of self-trapped states, 
we discuss another PL component induced mainly under excitation into 
the exciton band in each of 
PbF$_2$, PbCl$_2$, and PbBr$_2$. Figure \ref{fig5} shows  
the PL bands (solid lines) and the excitation spectra (dashed lines) 
of PbCl$_2$ and PbBr$_2$ below 8 K; 
both of them appear in the energy range higher than the BG-PL bands. 
In cubic PbF$_2$, such a PL band appears at about 4 eV. 
The PL components have been assigned to STE luminescence by many researchers 
\cite{Liidja,Kitaura,T.Fujita,Polak,Itoh,Kitaura2,Babin}. 
However, the stable range disagrees with that of self-trapped states 
in each crystal: In cubic PbF$_2$, the PL component at 4 eV is quenched 
around 30 K \cite{Liidja,Itoh} while the STHs are bleached around 175 K 
\cite{Nishi}. In PbCl$_2$, the PL component at 3.8 eV disappears at 25 K 
\cite{T.Fujita,Polak} while the STHs are stable up to 80 K 
and the STELs appear above 60 K \cite{Iwanaga4}. 
In PbBr$_2$, the PL component at 2.7 eV disappears at 30 K while the STHs 
become unstable at 20 K and the STEL are stable up to 120 K \cite{Iwanaga2}. 
In addition, the PL component is mainly induced under excitation into the 
exciton band while the STHs are induced also under interband excitation. 
Thus, all PL components attributed to STE luminescence show vital 
discrepancies with the STELs and STHs. 
If the PL components are intrinsic, the STEs responsible 
for the PL components need the different self-trapping centers from the 
STELs and STHs already detected with the ESR technique; 
however, further finding of the different self-trapping centers is quite 
improbable in these crystals. 

To consider the origin of the PL component, we point out that cubic PbF$_2$ 
is a super-ionic conductor \cite{Samara}, and PbCl$_2$ and PbBr$_2$ are 
high-ionic conductors \cite{Verwey,Ober}; 
therefore, they inevitably contain dense anion vacancies at higher than 
10$^{-17}$ cm$^{-3}$. In particular, lattice ions are absent with a 
high probability in the surface region because of the halogen desorption 
connected to the high-ionic conductivity. The vacancies may affect and trap 
the excitons induced by one-photon excitation only in the surface region. 
The excitation profiles of the PL components in the three crystals 
suggest that the excitons 
created by one-photon excitation behave similarly in the relaxation, and 
moreover the excitons are not coincident with any self-trapping center found 
experimentally to date. Therefore, the vacancy-associated relaxation of 
excitons seems to be a convincing explanation. At the end of discussion, 
we note that experimental data on cubic PbF$_2$ are partly inconsistent 
\cite{Liidja,Itoh}; the excitation profiles of PL component at 4 eV presented 
in the two references are different with each other. As for PbF$_2$, 
further study of luminescence and structures of 
localized states is necessary to clarify the exciton relaxation fully.

\section{Concluding remarks}
We have surveyed exciton relaxation in lead halides on the basis of 
recent experimental results. In PbBr$_2$ and PbCl$_2$, spontaneous 
exciton dissociation has been revealed with the luminescence spectroscopy 
and ESR technique. 
On the other hand, PbF$_2$ requires further investigation in view of 
luminescence properties and structures of localized states. 

In cubic PbF$_2$, PbCl$_2$, and PbBr$_2$, the PL components with similar 
excitation profiles are discussed from the 
correlation with the STELs and STHs, so that they are inconsistent with any 
STEL and STH. Consequently, they are unlikely to be intrinsic. To identify 
the origins, the examination with the optically detected magnetic-resonance 
technique is most preferable.

\begin{acknowledgments}
We would like to thank J.~Azuma, M.~Shirai, and K.~Tanaka for the 
collaboration in ESR experiments and the valuable discussion about exciton 
relaxation in lead halides.
\end{acknowledgments}

\end{document}